\documentclass[twocolumn,showpacs,amsmath,amssymb]{revtex4}
\input{epsf}

\usepackage{graphicx}
\usepackage{array}
\usepackage{longtable}
\usepackage{rotating,booktabs}
\usepackage{booktabs,threeparttable}
\usepackage{bm}

\begin{document}

\title{Tune-out wavelengths for potassium}

\author{Jun Jiang$^1$, Li-Yan Tang$^{1,2}$ and J. Mitroy$^{1}$}

\affiliation {$^{1}$School of Engineering, Charles Darwin 
University, Darwin NT 0909, Australia}

\affiliation {$^2$State Key Laboratory of Magnetic Resonance and
Atomic and Molecular Physics, Wuhan Institute of Physics and
Mathematics, Chinese Academy of Sciences, Wuhan 430071, P. R. China}

\date{\today}

\begin{abstract}
The five longest tune-out wavelengths for the potassium atom are
determined using a relativistic structure model which treats the 
atom as consisting of a single valence electron moving outside a  
closed shell core.  The importance of various terms in the dynamic 
polarizability in the vicinity of the $4p_J$, $5p_J$ and $6p_J$  
transitions are discussed. 

\end{abstract}

 \maketitle

\section{Introduction}

The dynamic polarizability of an atom gives a measure of the energy shift
of the atom when it is exposed to an electromagnetic field 
\cite{miller77a,mitroy10a}.  For an atom in any given
state, one can write
\begin{equation}
\Delta E \approx - \frac{1}{2} \alpha_d(\omega) F^2
\end{equation}
where $\alpha_d(\omega)$ is the polarizability of the quantum state at frequency
$\omega$, and $F$ is a measure of the strength of the AC electromagnetic field.
The limiting value of the dynamic polarizability in the $\omega \to 0$ limit is
the static dipole polarizability.

The dynamic polarizability will go to zero for certain frequencies of
the applied electromagnetic field.  The wavelengths at which the
polarizability goes to zero are called the tune-out wavelengths
\cite{leblanc07a,arora11a}.  Atoms trapped in an optical lattice
can be released by changing the wavelength of the trapping laser
to that of the tune-out wavelength for that atom.  Very recently,
tune-out wavelengths have been measured for the rubidium and the
potassium atoms \cite{holmgren12a,herold12a}.  The advantage of a tune-out 
wavelength measurement is that it effectively a null experiment, it 
measures the frequency
at which the polarizability is equal to zero.  Therefore it does not
rely on a precise determination of the strength of an electric field
or the intensity of a laser field.

In the present manuscript a calculation of the five longest tune-out
wavelengths for the potassium atom is presented.  The method applied
is a fully relativistic version of a non-relativistic semi-empirical 
fixed core approach that has been successfully applied to the description 
of many one and two electron atoms \cite{mitroy88d,mitroy03f,mitroy09a,mitroy09b}.  
An extensive discussion is made about those parts of the oscillator strength 
sum rule that have the largest influence in the determination of the tune-out 
wavelengths.      

\section{Formulation}

The calculation methodology is as follows.  The first step involves
a Dirac-Fock (DF) calculation of the K$^+$ ground state.  The single
particle orbitals are written as linear combinations of analytic
basis functions.  The functions chosen are the $S$-spinors introduced
by Grant and Quiney \cite{grant00a,grant07a}.  S-spinors can be regarded 
as relativistic generalizations of the familiar Slater type orbital.  

The effective interaction of the valence electrons with the core is then
written
\begin{eqnarray}
H  &=&  c\bm{\alpha}\cdot\bm{p}+\beta mc^2 + V_{\rm core}({\bf r})  \ , 
\end{eqnarray}
where $m$ is the electron mass, $c$ is the speed of light, $\bm{p}$ is
the momentum operator, $\bm{\alpha}$ and $\beta$ are $4\times 4$ matrices
of the Dirac operators \cite{grant07a}. The core operator is 
\begin{eqnarray}
V_{\rm core}({\bf r})  &=&  -\frac{Z}{r}+ V_{\rm dir}({\bf r}) + V_{\rm exc}({\bf r}) +  V_{\rm p}({\bf r}) \ .
\end{eqnarray}
The direct and exchange interactions of the valence electron with
the DF core were calculated exactly.  The $\ell$-dependent polarization
potential, $V_{\rm p}$, was semi-empirical in nature with the
functional form
\begin{equation}
V_{\rm p}(r)  =  -\sum_{\ell j} \frac{\alpha_{\rm core} g_{\ell j}^2(r)}{2 r^4}
 |\ell j \rangle \langle \ell j| .
 \label{polar1}
\end{equation}

The factor, $\alpha_{\rm core}$
is the static dipole polarizability of the core and
$g_{\ell j}^2(r) = 1-\exp\bigl(-r^6/\rho_{\ell j}^6 \bigr)$
is a cutoff function designed to make the polarization potential
finite at the origin.  The cutoff parameters, $\rho_{\ell j}$ were
tuned to reproduce the binding energies of the $ns$
ground state and the $np$, and $nd$ excited states and are 
listed in Table \ref{tab1}.

\begin{table}
\caption{\label{tab1} The cutoff parameters, $\rho_{\ell j}$ of the 
core polarization potential.}
\begin{ruledtabular}
\begin {tabular}{ccc}
  $\ell$      &   $J$ &  $\rho_{\ell j}$ ($a_0$) \\
\hline
$s$  &  1/2  &   2.1360    \\
$p$  &  1/2  &   2.0324    \\
     &  3/2  &   2.0289    \\
$d$  &  3/2  &   2.3610     \\
     &  5/2  &   2.3633     \\
\end{tabular}
\end{ruledtabular}
\end{table}

The effective Hamiltonian for the valence electron was diagonalized 
in a large $L$-spinor basis \cite{grant00a}.  $L$-spinors can be regarded 
as a relativistic generalization of the Laguerre type orbitals that are 
often used when solving the Schrodinger equation \cite{mitroy03f}.   
This basis can be enlarged towards completeness
without any linear dependence problems occurring.  There is effectively 
no error due to the incompleteness of the basis set in the present calculation.
The present relativistic configuration interaction plus core polarization 
calculations typically used 50 positive energy and 50 negative energy 
$L$-spinors for each $(\ell,j)$ symmetry.  This approach is named the 
relativistic configuration interaction plus core polarization 
approach (RCICP).  The non-relativistic approach from which the method 
is derived is called the configuration interaction plus core polarization 
(CICP) method \cite{mitroy03f}.  For the purpose of comparison, we present 
results of calculations using the all-order single-double implementation 
of relativistic many body perturbation theory \cite{safronova08a,safronova13a} 
(MBPT-SD).  The area of commonality between the RCICP and 
MBPT-SD approaches is that both calculations have minimal numerical 
uncertainties.  The two methods use different approximations to 
treat the interaction with the core and core-valence correlations, 
but the subsequent calculations within their respective theoretical 
frameworks have effectively no significant errors due to basis set 
incompleteness.  

\section{Results} 

\subsection{Energies} 

Table \ref{tab2} gives the energies of some of the low lying states
of potassium. One of the interesting aspects of the table concerns the
spin-orbit splitting of the $5p_J$ and $6p_J$ states.  The polarization
potential parameters $\rho_{1,1/2}$ and $\rho_{1,3/2}$
were tuned to give the correct
spin-orbit splitting of the $4p_J$ states.  Making this choice resulted in
the spin-orbit splittings for the $5p_J$ and $6p_J$ states also being
very close to experiment.  Similarly, tuning the $\rho_{2,3/2}$ and 
$\rho_{3,5/2}$ parameters to give the correct $3d_J$ spin-orbit splitting 
also resulted in the spin-orbit splittings for the $4d_J$ and $5d_J$ levels 
also in agreement with experiment.    

\begin{table}
\caption{\label{tab2} Theoretical and experimental energy levels (in Hartree)
for some of the low-lying states of K. The energies are given relative
to the energy of the K$^{+}$ core. The experimental data were taken from
the National Institute of Science and Technology (NIST) tabulation 
\cite{nistasd500}.}
\begin{ruledtabular}
\begin {tabular}{cccc}
          &  $J$   & Present        &  Experiment  \\
\hline
$4s$  &  1/2  &    $-$0.1595191  &      $-$0.1595165 \\
$4p$  &  1/2  &    $-$0.1003515  &      $-$0.1003516 \\
      &  3/2  &    $-$0.1000886  &      $-$0.1000886 \\
$5s$  &  1/2  &    $-$0.0636441  &      $-$0.0637124 \\
$3d$  &  5/2  &    $-$0.0613971  &      $-$0.0613972 \\
      &  3/2  &    $-$0.0613867  &      $-$0.0613867 \\
$5p$  &  1/2  &    $-$0.0469469  &      $-$0.0469687 \\
      &  3/2  &    $-$0.0468616  &      $-$0.0468832 \\
$4d$  &  5/2  &    $-$0.0346107  &      $-$0.0346862 \\
      &  3/2  &    $-$0.0346058  &      $-$0.0346813 \\
$6s$  &  1/2  &    $-$0.0344071  &      $-$0.0344418 \\
$6p$  &  1/2  &    $-$0.0273728  &      $-$0.0273861 \\
      &  3/2  &    $-$0.0273345  &      $-$0.0273476 \\
\end{tabular}
\end{ruledtabular}
\end{table}

\subsection{Dipole matrix elements} 

Table \ref{tab3} gives the reduced matrix elements for a number 
of the low-lying transitions of the potassium atom.  These matrix 
elements were computed with a modified transition operator 
\cite{hameed68a,hameed72a,mitroy88d}, 
e.g. 
\begin{equation}
{\bf r} = {\bf r} - \left(1 - \exp(-r^6/\rho^6) \right)^{1/2} \frac{\alpha_d {\bf r}}{r^3} 
\label{dipole}
\end{equation}
The cutoff parameter used in Eq.~(\ref{dipole}) was 2.176 $a_0$, 
the average of the $s$, $p$ and $d$ cutoff parameters (note, the
weighting of the $s$ was doubled to give it same weighting as the 
two $p$ and $d$ orbitals).  These matrix elements are compared to the 
MBPT-SD matrix elements \cite{safronova13a,safronova13c}.  
Some reduced matrix elements derived from experiment are also given 
\cite{safronova13a}.  It should be noted that there are some small 
differences between the most recent MBPT-SD reduced matrix elements 
\cite{safronova13a} and earlier work using this method 
\cite{safronova08b}.   

The agreement between the RCICP and MBPT-SD calculations 
for the larger matrix elements is very good.  The two calculations 
agree to better that 1$\%$ for the $4s-4p$, $4p-5s$, $5p-6s$ and $3d -4p$ 
transition manifolds.  The same level of agreement is achieved for the 
non-relativistic CICP matrix elements.  The CICP matrix 
elements are taken from earlier calculations of dispersion coefficients 
for atomic pair involving potassium \cite{mitroy03f,mitroy05b,mitroy03g,mitroy07d}.      
The agreement between the CICP, RCICP and MBPT-SD matrix is not 
so good for transitions with much smaller matrix elements 
such as $4s \to 6p_J$.      

\begin{table}
\caption{\label{tab3} Comparison of reduced dipole matrix elements 
(a.u.) for the principal transitions of potassium with experimental 
values and other calculations.}
\begin{ruledtabular}
\begin {tabular}{lcccc}
Transition              &   RCICP     &  MBPT-SD   & CICP  &  Experiment  \\
                        &             &  \cite{safronova99a,safronova08b,safronova13a,safronova13c}  &   \\
\hline
$4s$ - $4p_{1/2}$       &  4.1030     &   4.098     & 4.1017      &  4.102(5) \cite{volz96a} \\ 
                        &             &             &             &  4.106(4) \cite{falke06a} \\
$4s$ - $4p_{3/2}$       &  5.8016     &   5.794     & 5.8006      &  5.800(8) \cite{volz96a}   \\  
                        &             &             &             &    5.807(7) \cite{falke06a} \\
$4s$ - $5p_{1/2}$       &  0.2634     &   0.271(5)  & 0.2696                                   \\
$4s$ - $5p_{3/2}$       &  0.3886     &   0.398(8)  & 0.3812                                    \\
$4s$ - $6p_{1/2}$       &  0.0756     &   0.084     & 0.0791          \\
$4s$ - $6p_{3/2}$       &  0.1162     &   0.128     & 0.1118          \\
$4p_{1/2}$ -$5s$        &  3.8879     &   3.855(1)  & 3.9058           \\
$4p_{3/2}$ -$5s$        &  5.5384     &   5.54(1)   & 5.5236           \\
$5s$ - $5p_{1/2}$       &  9.4967     &   9.49(3)   & 9.4918          \\
$5s$ - $5p_{3/2}$       & 13.410      &  13.40(4)   & 13.423          \\
$5p_{1/2}$ -$6s$        &  8.7766     &   8.79(2)   & 8.8088           \\
$5p_{3/2}$ -$6s$        & 12.490      &   12.50(2)  & 12.458            \\
$4p_{1/2}$ -$3d_{3/2}$  &  7.9662     & 7.97(3)     & 7.9812      &   7.979(35)\cite{safronova13a} \\
$4p_{3/2}$ -$3d_{3/2}$  &  3.5731     & 3.57(1)     & 3.5693      &   3.578(16) \cite{safronova13a} \\
$4p_{3/2}$ -$3d_{5/2}$  & 10.719      &   10.72(3)  & 10.708      &  10.734(47) \cite{safronova13a} \\
$4p_{1/2}$ -$4d_{3/2}$  &  0.1403     &   0.1121(8) & 0.1246          \\
$4p_{3/2}$ -$4d_{3/2}$  &  0.0529     &   0.0400(1) & 0.0557          \\
$4p_{3/2}$ -$4d_{5/2}$  &  0.1553     &   0.1170(4) & 0.1672          \\
$3d_{3/2}$ -$5p_{1/2}$  &  7.1687     &   7.2(1)    & 7.1476             \\
$3d_{3/2}$ -$5p_{3/2}$  &  3.1934     &   3.19(9)   & 3.1965             \\
$3d_{5/2}$ -$5p_{3/2}$  &  9.5743     &   9.6(1)    & 9.5895             \\
$5p_{1/2}$ -$4d_{3/2}$  & 17.040      &   17.04(6)  & 17.076            \\
$5p_{3/2}$ -$4d_{3/2}$  &  7.6432     &   7.64(3)   & 7.6367             \\
$5p_{3/2}$ -$4d_{5/2}$  & 22.932     &   22.93(8)  & 22.910            \\
$\frac{|\langle 4p_{3/2}\|D\|4s\rangle|^2}{|\langle 4p_{1/2}\|D\|4s\rangle|^2}$ & 1.99935 & 1.9987 \cite{holmgren12a} & 2.0 & 2.0005(40) \cite{holmgren12a} \\ 
  &    &     &  &  1.999(4) \cite{volz96a} \\  
  &    &     &  &  2.002(3) \cite{falke06a} \\
 &   &    &  & 2.01  \cite{shabanova84a} \\
$\frac{|\langle 5p_{3/2}\|D\|4s\rangle|^2}{|\langle 5p_{1/2}\|D\|4s\rangle|^2}$ & 2.17787 & 2.17964 & 2.0 & 2.15 \cite{shabanova84a} \\
$\frac{|\langle 6p_{3/2}\|D\|4s\rangle|^2}{|\langle 6p_{1/2}\|D\|4s\rangle|^2}$ & 2.35799 & 2.31894  & 2.0 & 2.28 \cite{shabanova84a} \\
\end{tabular}
\end{ruledtabular}
\end{table}

The ratio of the line strengths for the $4s \to 4p_J$ transition 
would be exactly 2.0 in a non-relativistic model.  Experiment and 
theory both indicate that the difference of the $4s - 4p$ 
transition ratio from 2.0 is very small.  The current calculation 
and the MBPT-SD calculation indicate that the matrix element 
ratio is slight smaller than 2.0.  
However, the ratio differs substantially from 2.0 for the $4s \to 5p_J$ 
and $4s \to 6p_J$ transitions.  The
main cause for the deviation of the ratio from 2.0 is the
slightly different wavefunction for the spin-orbit doublet arising 
from the slightly different energies \cite{migdalek98a}.  We
have done non-relativistic calculations and have been able to
reproduce the ratios given in Table \ref{tab3} by simply tuning 
the polarization potential to separately give the experimental 
binding energies of each spin-orbit doublet.  

\subsection{Polarizabilities and tune-out wavelengths} 

The computations of the
static polarizabilities utilized the RCICP matrix elements but
with the excitation energies for the $4p_J$, $5p_J$ and $6p_J$ set
to be those of experiment.  The dynamic polarizability is defined 
\begin{eqnarray}
\alpha(\omega) & = & \sum_{n} \frac {f_{0n} } {(\epsilon_{0n}^2-\omega^2)} \ , 
\label{alpha1}
\end{eqnarray}
where $f_{0n}$ is the oscillator strength for the dipole transition.
For low frequencies, the dynamic polarizability can be expanded   
\begin{eqnarray}
\alpha(\omega) & = & \alpha(0) + S(-4)\omega^2 + \ldots \ .
\label{alpha2}
\end{eqnarray}
where $\alpha(0)$ is the static dipole polarizability calculated at 
$\omega = 0$ and $S(-4)$ is calculated using the oscillator strength 
sum rule,    
\begin{eqnarray}
S(-4) & = &  \sum_{n} \frac {f_{0n} } {\epsilon_{0n}^4} \ .
\label{alpha3}
\end{eqnarray}
Polarizabilities for the potassium ground state 
from different sources are listed in Table \ref{tab4}.  The 
present RCICP calculation gave essentially the same 
polarizability, 290.1 $a_0^3$ as its non-relativistic CICP 
predecessor.  The non-relativistic CICP model had its valence 
energies tuned to experimental energies (in this case, the 
$(2J+1)$ weighted energy of any spin-orbit doublet), just like the present 
calculation. The $4s - 4p_J$ radial matrix elements are dominated 
by long range form of the wave function, and that is practically 
the same for the CICP and the relativistic RCICP calculations.    
The present RCICP calculation agrees with experiment 
\cite{derevianko99a,holmgren10a} within the experimental 
uncertainties.  

The relativistic coupled cluster calculation 
(RCCSD) \cite{lim99a} gives a dipole polarizability that is about 
3$\%$ larger than experiment.    

\begin{table}
\caption{\label{tab4} Static dipole polarizabilities (in a.u.) for potassium. A
short description of the details behind some of the polarizabilities can 
be found in Ref.~\cite{mitroy10a}.}
\begin{ruledtabular}
\begin {tabular}{lcc}
                                         &   $\alpha_1$    &  $\alpha_2$ $(10^3)$       \\ \hline 
Theory: Present RCICP                    &  290.1   &  5.000    \\
Theory: CICP \cite{mitroy03f}            &  290.0   &  5.005    \\
Theory: MBPT-SD   \cite{safronova08b}    &  289.3   &  5.018    \\
Theory: RCCSD      \cite{lim99a}         &  301.28   &  5.018       \\
Expt. $E \times H$    \cite{molof74a}    &  293(6)   &            \\
Expt: Interferometer  \cite{holmgren10a} &  290.8(1.4)   &             \\
Hybrid: Sum rule  \cite{derevianko99a}   &  290.2(8)   &             \\  
\end{tabular}
\end{ruledtabular}
\end{table}

The core polarizability is given by a pseudo-oscillator strength 
distribution \cite{margoliash78a,kumar85a,mitroy03f}.  The distribution 
is tabulated in Table \ref{tab5}.  The distribution is derived from 
the single particle energies of the Hartree-Fock core.  Each separate 
$(n,\ell)$ level is identified with one transition with a pseudo-oscillator 
strength equal to the number of electrons in the shell.  The excitation
energy is set by adding a constant to the Koopman energies and 
tuning the constant until the core polarizability from the 
oscillator strength sum rule is equal to the know core polarizability.  
In the present case, the K$^+$ core polarizability was set to 5.47 a.u.
\cite{opik67a,mitroy93a}. 

\begin{table}
\caption{\label{tab5} Pseudo-spectral oscillator strength distribution for the 
potassium core.   Energies are given in a.u..  }
\begin{ruledtabular}
\begin {tabular}{lcc}
$n$ &    $\varepsilon_n$   & $f_n$  \\
\hline
1 &   133.6890020  &    2.0  \\ 
2 &    14.6459330  &    2.0  \\ 
3 &    11.6752580  &    6.0  \\ 
4 &    1.90477720  &    2.0  \\ 
5 &    1.11041710  &    6.0  \\ 
\end{tabular}
\end{ruledtabular}
\end{table}

The tune-out wavelengths all tend to be close to the wavelengths 
for excitation of the $np_J$ excited states.   There are two scenarios 
that lead to tune out wavelengths.  In the first, the tune-out wavelength 
occurs in the middle of an $np_J$ spin-orbit doublet. The wavelength will 
be shorter than the transition wavelength to the $np_{1/2}$ state and longer than 
the wavelength to the $np_{3/2}$ state.  When this occurs, the dynamic
polarizabilities of the $np_{1/2}$ and $np_{3/2}$ states will have 
the opposite sign and this will lead to a zero in the total dynamic 
polarizability occurring for all spin-orbit doublets.   
The second scenario leading to a tune-out wavelength 
occurs when the wavelength is shorter than that for excitation of the 
$4p_J$ states.  When this occurs, the contribution to the dynamic 
polarizability from the $4p_J$ states become negative.  This leads to a 
series of tune-out wavelengths occurring just below the excitation  
energies of the $5p_{1/2}$, $6p_{1/2}$, $7p_{1/2}$, \ldots states.   

\begin{table*}
\caption{\label{tab6} Breakdown of contributions to the potassium ground state
polarizability at different wavelengths. }
\begin{ruledtabular}
\begin {tabular}{lcccccc}
 $\lambda$ (nm)      &   $\infty$ & 768.97075         & 405.9173        & 404.7217            & 344.9099   & 344.7861      \\
 $\omega$ (a.u.)      &     0     & 0.059252386       & 0.11224787       & 0.11257945         &0.13210218   & 0.13214964   \\
\hline  
$4s - 4p_{1/2}$      &   94.8454  &$-$32032.4796      & $-$36.4876       &  $-$36.1911        & $-$23.7987  & $-$23.7774    \\
$4s - 4p_{3/2}$      &  188.7902  &   32025.7787      & $-$73.5278       &  $-$72.9282        & $-$47.9006  & $-$47.8575    \\
$4s - 5p_{1/2}$      &    0.2054  &       0.2842      &    38.5906       & $-$364.9459        &  $-$0.5438  &  $-$0.5424    \\
$4s - 5p_{3/2}$      &    0.4469  &       0.6180      &    65.4323       &    468.0690        &  $-$1.1901  &  $-$1.1870    \\
$4s - 6p_{1/2}$      &    0.0144  &       0.0181      &     0.0519       &      0.0527        &   33.7885   & $-$49.5664    \\
$4s - 6p_{3/2}$      &    0.0340  &       0.0426      &     0.1221       &      0.1240        &   33.7302   &   117.0156    \\
Remainder Valence    &    0.2426  &       0.2528      &     0.2957       &      0.2963        &   0.3712    &     0.3717    \\
$\alpha_{\rm core}$  &    5.4708  &       5.4852      &     5.5229       &      5.5232        &   5.5433    &     5.5434    \\
Total                &  290.050   &             0     &  0                &   0               &  0          &          0  \\  
\end{tabular}
\end{ruledtabular}
\end{table*}

Simplified expressions can be used to describe the dynamic polarizabilities 
on the vicinity of the tune-out wavelengths.  The first tune-out 
wavelength occurs when the wavelength lies between the $4p_{1/2}$ and 
$4p_{3/2}$ resonant wavelengths.  The dynamic polarizability here 
can be written,   
\begin{equation}
\alpha_1({\omega}) = \frac{f_{4p_{1/2}}} { (\Delta E_{\rm 4p_{1/2}}^2 - \omega^2 ) }
                + \frac{f_{4p_{3/2}}} { (\Delta E_{\rm 4p_{3/2}}^2 - \omega^2 ) }
                + \alpha_{\rm rem}(\omega)\,,
\end{equation}
where $\alpha_{\rm rem}(\omega)$ is remainder part of dynamic dipole polarizability.  
The energy difference, $\Delta E_{\rm 4p_{3/2}}$ can be parameterized 
as $\Delta E_{\rm 4p_{3/2}}=  \Delta E_{\rm 4p_{1/2}}(1 + \delta)$.  Parameterizing 
the line strength, $S$ (the square of the reduced matrix element), as
\begin{eqnarray}
S(4s\to 4p_{3/2}) =  S(4s \to 4p_{1/2}) (2 + R)
\end{eqnarray}
leads to
\begin{eqnarray}
\alpha_1(\omega)& = & \frac{f_{4p_{1/2}}} { (\Delta E_{\rm 4p_{1/2}}^2 - \omega^2 ) }
                + \frac{f_{4p_{1/2}}(2+R)(1+\delta)} { [(\Delta E_{\rm 4p_{1/2}})^2(1+\delta)^2 - \omega^2 ] } \nonumber \\  
               & + & \alpha_{\rm rem}(\omega)\ .
\label{exact}
\end{eqnarray}
The dipole oscillator strength $f_{4p_{1/2}}$ is obtained by multiplying the
reduced matrix element, with the experimental $4s \to 4p_{1/2}$ energy difference.  
The value of $R$ is simply the ratio of computed line strength coming from 
the RCICP calculations.   It can be seen from Table \ref{tab6} that 
the remainder polarizability, $\alpha_{\rm rem}(\omega)$ only makes a small 
contribution to the total polarizability.  The remainder polarizability varies 
relatively slowly with wavelength in the vicinity of the tune-out wavelength.       

Table \ref{tab7} illustrates the variation in the tune-out wavelength 
with respect to variations in $R$ and $\alpha_{\rm rem}(\omega)$.  
The contributions to the 
polarizability from the $4p_{1/2}$ and $4p_{3/2}$ transitions 
are 5,000 times larger than those from every other transition.    
A change in $\alpha_{\rm rem}(\omega)$ of 1.0 a.u leads to the tune-out 
wavelength changing by 0.00002 nm.  The tune-out wavelength is 
much more sensitive to variations in $R$.  A change in $R$ to $-$0.005 
leads to the tune-out wavelength changing by 0.0016 nm.  A value of 
$R = -0.005$ is 3 times larger than the MBPT-SD value and is seven times larger 
than the RCICP value of $R$.   

A different parameterization should be used in the vicinity of the 
of the excited states with $n > 4$.     
\begin{eqnarray}
\alpha_1({\omega}) &= & \alpha_{4p}(\omega) +  \alpha_{\rm rem}(\omega)    
       +  \frac{f_{np_{1/2}}} { (\Delta E_{\rm np_{1/2}}^2 - \omega^2 ) } \nonumber \\  
     &+& \frac{f_{np_{1/2}}(2+R)(1+\delta)} { [(\Delta E_{\rm np_{1/2}})^2(1+\delta)^2 - \omega^2 ] }   
\end{eqnarray} 
Here the polarizability arising from the $4s \to 4p_J$ transitions is 
retained as a separate term since it is much larger than the remainder.

\begin{table}[tbh]
\caption{Values of the tune-out wavelength for the K atom. The experimental 
transition wavelengths are taken from \cite{sansonetti11a,radziemski95a}.  The first entry 
lists the tune-out wavelength as computed with RCICP matrix elements.  The 
other entries exhibit the changes to the tune-out wavelengths when changes 
are made to matrix elements underlying the oscillator strength sums.} 
\label{tab7}
\begin{ruledtabular}
\begin{tabular}{lll}
\multicolumn{1}{l}{ Resonance} & \multicolumn{1}{l}{$\omega$ (a.u.) }  & \multicolumn{1}{l}{$\lambda$ (nm)} \\
\hline
$\Delta E_{4s-4p_{1/2}}$                                       & 0.059164859   & 770.10836  \\
$\Delta E_{4p_{3/2}}$                                       & 0.059427807   & 766.70089 \\
$R=-0.0006476$, $\alpha_{\rm rem}(\omega) = 6.701$          & 0.059252386   & {\bf 768.97077} \\
$\alpha_{\rm rem}(\omega) = 5.701$                          & 0.059252387   & 768.97075 \\
$R = -0.005 $                                               & 0.059252513   & 768.96912 \\
MBPT-SD \cite{arora11a}                                     & 0.0592524(2)  & 768.971(3) \\
Experiment \cite{holmgren12a}                               & 0.0592523(1)  & 768.9712(15) \\
\hline
$\Delta E_{4s-5p_{1/2}}$                                       & 0.11254778    & 404.8356    \\
$\Delta E_{4s-5p_{3/2}}$                                       & 0.11263324    & 404.5285 \\
$R=0.17787$, $\alpha_{\rm rem}(\omega) =-104.023 $          & 0.11224787    & {\bf 405.9173} \\
$R=0.15787$                                                 & 0.1122499     & 405.9100  \\
$|\langle 5p_J\|D\|4s\rangle| \times 1.03$                  & 0.1122120     & 405.9943 \\ 
MBPT-SD \cite{arora11a}                                     & 0.11223(1)    & 405.98(4) \\
\hline
$R=0.17787$, $\alpha_{\rm rem}(\omega) =-103.300$           & 0.11257946   & {\bf 404.7217} \\
$R=0.15787$,                                                & 0.11257967   & 404.7210 \\
$|\langle 5p_J\|D\|4s\rangle| \times 1.03$                  & 0.11257898   & 404.7228 \\ 
MBPT-SD \cite{arora11a}                                     & 0.11258(1)   & 404.72(4) \\
\hline 
$\Delta E_{4s-6p_{1/2}}$                                       & 0.13213040     & 344.8363  \\
$\Delta E_{4s-6p_{3/2}}$                                       & 0.13216885     & 344.7359  \\
$R=0.35799$, $\alpha_{\rm rem}(\omega) =-67.5188$           & 0.13210218     & {\bf 344.9099}   \\
$R=0.30799$,                                                & 0.13210256     & 344.9089    \\
$|\langle 6p_J\|D\|4s\rangle| \times 1.10$                  & 0.13209360     & 344.9323 \\ 
MBPT-SD \cite{arora11a}                                     & 0.1320933(4)   & 344.933(1) \\
\hline
$R=0.35799$, $\alpha_{\rm rem}(\omega) =-67.4492$           & 0.13214964    & {\bf 344.7861}  \\
$R=0.30799$,                                                & 0.13214992    & 344.7853  \\
$|\langle 6p_J\|D\|4s\rangle| \times 1.10$                  & 0.13214825    & 344.7897  \\ 
\end{tabular}
\end{ruledtabular}
\end{table}

The tune-out wavelengths in the vicinity of the $5p_J$ levels illustrate 
clearly how the $\alpha_{\rm rem}(\omega)$ and $R$ are of different 
importance depending on whether the tune-out energy is located between 
the $5p_{1/2}$ and $5p_{3/2}$ levels or before the $5p_{1/2}$ level.   

Table \ref{tab6} shows that the tune-out wavelength at energies below the $5p_{1/2}$ 
excitation threshold is caused by the cancellation of the $5p_J$ and core 
contributions with those of coming from the $4p_J$ levels.  This tune-out 
wavelength is effectively determined by ratio of the $4p$ and $5p$ 
oscillator strengths.  The contribution of the core is small in absolute 
terms, and a 5$\%$ uncertainty in the core polarizability will have a 
small effect on the tune-out wavelength.  The tune-out wavelength is 
predominantly determined by the relative size of the $4s \to 4p_J$ and 
$4s \to 5p_J$ matrix elements.  This can be seen from Table \ref{tab7}.   
Increasing the $4s \to 5p_J$ matrix elements by 3$\%$ (roughly the 
difference with the MBPT-SD matrix elements) leads to the tune-out
wavelength increasing by 0.077 nm.  Changing the value of $R$, by 
0.02 leads to the tune-out wavelength changing by 0.007 nm.   
It should be noted that our definition of $R$ does imply an
overall increase in the total oscillator strength to the 
$5p_J$ states.     

The tune-out wavelength at the energies between the $5p_{1/2}$ and 
$5p_{3/2}$ levels does have some dependence on $\alpha_{rem}$ since 
it now incorporates the contribution from the $4s \to 4p_J$ transitions.    
A 5$\%$ change in the matrix element leads to a change in the tune-out 
wavelength of 0.0024 nm, this is 60 times smaller than the effect on 
the tune-out wavelength at 405.9173 nm.  The sensitivity to a change 
in the value of $R$ by 0.02 was only 0.0007 nm.  In relative terms,  
the tune-out wavelength is more sensitive to the value of $R$ than 
$|\langle 5p_J\|D\|4s\rangle|$ in the $5p_{1/2}$ to $5p_{3/2}$ energy 
gap, than it is in the energy region before $5p_{1/2}$ excitation.  

Table \ref{tab7} also gives the tune-out wavelengths in the vicinity 
of the $6p_J$ levels.  Once again the tune-out wavelengths are very sensitive 
to the absolute size of the $|\langle 6p_J\|D\|4s\rangle|$ transition 
matrix element.   A 10$\%$ change in the matrix element (the difference 
between the RCICP and MBPT-SD calculations) leads to a change of 0.0234 nm 
in the tune-out wavelength just below the $6p_{1/2}$ threshold.  The 
considerations that determine the values of the tune-out wavelengths 
in the vicinity of the $5p_J$ states also apply to the tune-out 
wavelengths in the vicinity of the $6p_J$ states.  

Finally, it is noted that the tune-out wavelengths were also evaluated using 
the MBPT-SD matrix elements in Table \ref{tab3} but with other aspects 
of the calculation taken from the RCICP calculation.  The resulting tune-out 
wavelengths were identical to the MBPT-SD tune-out wavelengths in Table 
\ref{tab7} to all quoted digits.  

%------------------------------------------------------------------------------------------------------------------------------

%\begin{figure}[tbh]
%\centering{
%\includegraphics[width=8.4cm]{F2.eps}
%} \caption{ \label{fig2} (color online). Plot of $10^9 Z^2 \Delta
%\alpha_1$ as a function of nuclear charge, $Z$. }
%\end{figure}

\subsection{Uncertainties}

Part of this manuscript is focussed on the prediction of the tune-out 
wavelengths, but another and possibly more important part concerns 
the extraction of useful atomic structure information from   
an experimental value of the tune-out wavelength.  Knowledge of the 
$np_J$ tune-out wavelengths permits the determination of the 
$np_J$ oscillator strengths to a high degree of precision.   

The most important atomic parameters that contribute to the long 
wavelength dynamic polarizability are listed in Table \ref{tab8}.   
These parameters are derived from the RCICP calculations and 
uncertainties are estimated by examination of the difference 
with experiment or MBPT-SD calculations.   

\begin{table}
\caption{\label{tab8} Tabulation of atomic parameters, with estimated 
uncertainties that can be used reduced dipole matrix elements 
(a.u.) for the principal transitions of potassium with experimental 
values and other calculations.}
\begin{ruledtabular}
\begin {tabular}{lc}
Parameter              &   RCICP        \\ \hline 
$4s$ - $4p_{1/2}$      &  4.1030(7)   \\ 
$4s$ - $4p_{3/2}$      &  5.8016(10)    \\  
$4s$ - $5p_{1/2}$      &  0.2634(8)   \\
$4s$ - $5p_{3/2}$      &  0.3886(100)    \\
$4s$ - $6p_{1/2}$      &  0.0756(100)    \\
$4s$ - $6p_{3/2}$      &  0.1162(14)     \\
$\alpha_{\rm core}$  &    5.4708(1000)    \\
$S_{\rm core}(-4)$  &    4.10(80)   \\
$\alpha_{\rm core-valence}$  \cite{safronova99a} &   $-$0.13      \\
$\alpha_{\rm Remainder Valence}$  &   0.243(111)   \\
$S_{\rm Remainder Valence}(-4)$   &   2.67(134)   \\
\end{tabular}
\end{ruledtabular}
\end{table}

The determination of 
the $4p_{1/2}:4p_{3/2}$ line strength ratio is only weakly dependent 
on the value of the non $4p$ terms in the dynamic polarizability since 
these terms are small.    

Knowledge of the $np_J$ tune-out wavelengths permits the determination of the 
$np_J$ oscillator strengths for $n > 5$ to a high degree of precision.   The 
polarizability becomes zero when the contributions to the polarizability from the 
$np_J$ levels and the remainder cancel exactly.  The biggest terms in the 
remainder are $\alpha_{4p_J}$ and $\alpha_{\rm core}$ which together 
constitute 99$\%$ of the remainder.  Moreover, both of these terms 
are known with a reasonable degree of precision.  

The uncertainties in the experimental $4s \to 4p_J$ line strengths do not 
exceed 0.3$\%$ and the RCICP transition matrix element lies between two 
experimental estimates \cite{volz96a,falke06a}.  There is 2$\%$ 
variation between the RCICP and MBPT-SD estimates of the $4s \to 5p_J$ 
matrix elements and a 10$\%$ variation between $4s \to 6p_J$ matrix 
elements.  Taken together, $4s \to 5p_J$ and $4s \to 6p_J$ would 
contribute less than 0.01 a.u. to the uncertainty of the total polarizability 
at $\omega = 0$ a.u.   

The uncertainty in the core polarizability of 5.47 a.u.  
itself is stated to be about $2\%$ \cite{opik67a}.  This polarizability was  
based on the binding energies of the $4f$, $6f$ and $9f$ levels of 
potassium using spectral data from 1955 \cite{risberg56b}.  There 
is scope for an improvement in the precision of the core polarizability 
and this could be easily accomplished by spectroscopic experiments that 
measured the energies of the $ng$ levels.  The construction of pseudo-oscillator 
strength distribution for the $K^+$ core permits the energy variation 
of $\alpha_{\rm core}$ to be incorporated into the calculation.  The 
uncertainty of $20\%$ was estimated by using the same procedure to 
construct the pseudo-oscillator strength distribution to argon and 
making reference to a highly accurate pseudo-oscillator strength distribution  
\cite{kumar85a}.  

The core-valence term, $\alpha_{\rm core-valence}$ is a term that compensates 
for Pauli-Principle violating excitations from the core to the valence $4s$ 
orbital.  This value is sourced from an MBPT-SD calculation \cite{safronova99a} 
since it is not incorporated in the RCICP calculation.  No uncertainty has   
been assigned to this contribution to the polarizability.    

The valence remainder term contains contributions from highly excited discrete 
transitions as well as contributions from the continuum.  The present value is 
0.243 a.u. at $\omega = 0$.  This is more than twice the size of a MBPT-SD estimate  
of 0.07  a.u. \cite{safronova99a}.  However, much of the MBPT-SD valence remainder 
is computed in the DF approximation.  We have 
performed calculations in the DF approximation, and the DF oscillator 
strengths embedded in the continuum beyond the cooper minimum at 
0.010 a.u. \cite{zatsarinny10a} are typically a factor of 3 and 4 smaller than 
the RCICP oscillator strengths at those energies  Nevertheless, the 
uncertainty in this term has been conservatively assessed at 50$\%$.   
Comparisons of the RCICP and MBPT-SD oscillator strength distributions in the 
continuum,  and further comparisons with experimental photo-ionization cross 
sections would be helpful in refining the estimate and uncertainty of this 
rather small term.  

The parameter set and error budget in Table \ref{tab8} can be utilized to help convert 
tune-out wavelengths into oscillator strengths.  There is room for improvement 
in the parameter set.  Measurements of the tune-out wavelengths near the 
$5p_J$ excitation will result in better estimates of the $5p_J$ matrix 
elements.  Parameters obtained from theory do not have to be exclusively 
obtained from a single calculation, for example some information from Table 
\ref{tab8} might be best obtained from a MBPT-SD calculation while others,  
e.g. the valence remainder might be best estimated from the RCICP calculation 
or some RCICP/MBPT-SD hybrid.   

\section{Conclusion}
    
The five lowest tune-out wavelengths for the potassium atom are 
computed by a relativistic structure model.  A detailed analysis  
is performed regarding the contribution that the different terms 
make to the polarizability.  The results illustrate the 
dependence of the tune-out wavelengths on a relatively small number 
of atomic parameters.   
   
The lowest energy tune-out wavelength is primarily determined by 
the ratio of the line strengths for the $4s \to 4p_J$ transitions.    
The present calculation, and the MBPT-SD calculation are in 
agreement with existing experimental data \cite{holmgren12a}. 
The precision of the experiment would need to improve by an 
order of magnitude to provide a stringent test of the $4p_J$ 
states line strength ratio.  However, Holmgren {\em et. al.} 
\cite{holmgren12a} suggest that it might be possible to improve 
the precision by up to 3 orders of magnitude.   

The tune-out wavelengths near the $5p_J$ excited states are 
most sensitive to the ratio of the $4s \to 4p_J$ and  
$4s \to 5p_J$ matrix elements.  The remainder term 
incorporating all transitions except for the $4s \to 5p_J$ 
is dominated by the $4s \to 4p_J$ and core polarizabilities 
which comprise 99.5$\%$ of the remainder polarizability.   
The tune-out wavelengths here provide a means to determine 
the $5s \to 5p_J$ oscillator strengths to high precision.   
Measurement of the tune-out wavelength to a precision of 
0.01 nm would lead to oscillator strengths with a precision 
better than 1$\%$.    

Holmgren {\em et. al.} \cite{holmgren12a} suggested that measurements 
of the tune-out wavelengths near the $5p_{1/2}$ excited states could 
be used to determine the core polarizability.  We do not agree with this 
statement.  The remainder terms near the $5p_J$ excitation are dominated 
by $\alpha_{4p}(\omega)$ and $\alpha_{\rm core}(\omega)$.  The uncertainty 
in $\alpha_{4p}(\omega)$ at 405 nm would be about 0.3 a.u. and this 
uncertainty limits the precision with which the core polarizability 
could be measured. The preferred approach is to treat the 
$\alpha_{4p}(\omega)$ and $\alpha_{\rm core}(\omega)$ polarizabilities 
as known quantities with relatively small uncertainties and use measurements 
of the tune-out wavelength to extract precision values of the $5p_J$ 
oscillator strengths.    

This research was supported by the Australian Research Council
Discovery Project DP-1092620.  We thank Dr M S Safronova for helpful 
communications regarding her matrix elements. 

%%%%%%%%%%%%%%%%%%%%%%%% begin thebibliography  %%%%%%%%%%%%%%%%%%%%%%%%%%

%\bibliography{positron}

\begin{thebibliography}{39}
\expandafter\ifx\csname natexlab\endcsname\relax\def\natexlab#1{#1}\fi
\expandafter\ifx\csname bibnamefont\endcsname\relax
  \def\bibnamefont#1{#1}\fi
\expandafter\ifx\csname bibfnamefont\endcsname\relax
  \def\bibfnamefont#1{#1}\fi
\expandafter\ifx\csname citenamefont\endcsname\relax
  \def\citenamefont#1{#1}\fi
\expandafter\ifx\csname url\endcsname\relax
  \def\url#1{\texttt{#1}}\fi
\expandafter\ifx\csname urlprefix\endcsname\relax\def\urlprefix{URL }\fi
\providecommand{\bibinfo}[2]{#2}
\providecommand{\eprint}[2][]{\url{#2}}

\bibitem[{\citenamefont{Miller and Bederson}(1977)}]{miller77a}
\bibinfo{author}{\bibfnamefont{T.~M.} \bibnamefont{Miller}} \bibnamefont{and}
  \bibinfo{author}{\bibfnamefont{B.}~\bibnamefont{Bederson}},
  \bibinfo{journal}{Adv.~At.~Mol.~Phys.} \textbf{\bibinfo{volume}{13}},
  \bibinfo{pages}{1} (\bibinfo{year}{1977}).

\bibitem[{\citenamefont{Mitroy et~al.}(2010)\citenamefont{Mitroy, Safronova,
  and Clark}}]{mitroy10a}
\bibinfo{author}{\bibfnamefont{J.}~\bibnamefont{Mitroy}},
  \bibinfo{author}{\bibfnamefont{M.~S.} \bibnamefont{Safronova}},
  \bibnamefont{and} \bibinfo{author}{\bibfnamefont{C.~W.} \bibnamefont{Clark}},
  \bibinfo{journal}{J.~Phys.~B} \textbf{\bibinfo{volume}{43}},
  \bibinfo{pages}{202001} (\bibinfo{year}{2010}).

\bibitem[{\citenamefont{{Leblanc} and {Thywissen}}(2007)}]{leblanc07a}
\bibinfo{author}{\bibfnamefont{L.~J.} \bibnamefont{{Leblanc}}}
  \bibnamefont{and} \bibinfo{author}{\bibfnamefont{J.~H.}
  \bibnamefont{{Thywissen}}}, \bibinfo{journal}{\pra}
  \textbf{\bibinfo{volume}{75}}, \bibinfo{eid}{053612} (\bibinfo{year}{2007}).

\bibitem[{\citenamefont{{Arora} et~al.}(2011)\citenamefont{{Arora},
  {Safronova}, and {Clark}}}]{arora11a}
\bibinfo{author}{\bibfnamefont{B.}~\bibnamefont{{Arora}}},
  \bibinfo{author}{\bibfnamefont{M.~S.} \bibnamefont{{Safronova}}},
  \bibnamefont{and} \bibinfo{author}{\bibfnamefont{C.~W.}
  \bibnamefont{{Clark}}}, \bibinfo{journal}{\pra}
  \textbf{\bibinfo{volume}{84}}, \bibinfo{eid}{043401} (\bibinfo{year}{2011}).

\bibitem[{\citenamefont{{Holmgren} et~al.}(2012)\citenamefont{{Holmgren},
  {Trubko}, {Hromada}, and {Cronin}}}]{holmgren12a}
\bibinfo{author}{\bibfnamefont{W.~F.} \bibnamefont{{Holmgren}}},
  \bibinfo{author}{\bibfnamefont{R.}~\bibnamefont{{Trubko}}},
  \bibinfo{author}{\bibfnamefont{I.}~\bibnamefont{{Hromada}}},
  \bibnamefont{and} \bibinfo{author}{\bibfnamefont{A.~D.}
  \bibnamefont{{Cronin}}}, \bibinfo{journal}{Phys.~Rev.~Lett.}
  \textbf{\bibinfo{volume}{109}}, \bibinfo{eid}{243004} (\bibinfo{year}{2012}).

\bibitem[{\citenamefont{{Herold} et~al.}(2012)\citenamefont{{Herold}, {Vaidya},
  {Li}, {Rolston}, {Porto}, and {Safronova}}}]{herold12a}
\bibinfo{author}{\bibfnamefont{C.~D.} \bibnamefont{{Herold}}},
  \bibinfo{author}{\bibfnamefont{V.~D.} \bibnamefont{{Vaidya}}},
  \bibinfo{author}{\bibfnamefont{X.}~\bibnamefont{{Li}}},
  \bibinfo{author}{\bibfnamefont{S.~L.} \bibnamefont{{Rolston}}},
  \bibinfo{author}{\bibfnamefont{J.~V.} \bibnamefont{{Porto}}},
  \bibnamefont{and} \bibinfo{author}{\bibfnamefont{M.~S.}
  \bibnamefont{{Safronova}}}, \bibinfo{journal}{Phys.~Rev.~Lett.}
  \textbf{\bibinfo{volume}{109}}, \bibinfo{eid}{243003} (\bibinfo{year}{2012}).

\bibitem[{\citenamefont{Mitroy and Bromley}(2003{\natexlab{a}})}]{mitroy03f}
\bibinfo{author}{\bibfnamefont{J.}~\bibnamefont{Mitroy}} \bibnamefont{and}
  \bibinfo{author}{\bibfnamefont{M.~W.~J.} \bibnamefont{Bromley}},
  \bibinfo{journal}{Phys.~Rev.~A} \textbf{\bibinfo{volume}{68}},
  \bibinfo{pages}{052714} (\bibinfo{year}{2003}{\natexlab{a}}).

\bibitem[{\citenamefont{Mitroy et~al.}(1988)\citenamefont{Mitroy, Griffin,
  Norcross, and Pindzola}}]{mitroy88d}
\bibinfo{author}{\bibfnamefont{J.}~\bibnamefont{Mitroy}},
  \bibinfo{author}{\bibfnamefont{D.~C.} \bibnamefont{Griffin}},
  \bibinfo{author}{\bibfnamefont{D.~W.} \bibnamefont{Norcross}},
  \bibnamefont{and} \bibinfo{author}{\bibfnamefont{M.~S.}
  \bibnamefont{Pindzola}}, \bibinfo{journal}{Phys.~Rev.~A}
  \textbf{\bibinfo{volume}{38}}, \bibinfo{pages}{3339} (\bibinfo{year}{1988}).

\bibitem[{\citenamefont{Mitroy and Safronova}(2009)}]{mitroy09a}
\bibinfo{author}{\bibfnamefont{J.}~\bibnamefont{Mitroy}} \bibnamefont{and}
  \bibinfo{author}{\bibfnamefont{M.~S.} \bibnamefont{Safronova}},
  \bibinfo{journal}{Phys.~Rev.~A} \textbf{\bibinfo{volume}{79}},
  \bibinfo{pages}{012513} (\bibinfo{year}{2009}).

\bibitem[{\citenamefont{Mitroy et~al.}(2009)\citenamefont{Mitroy, Zhang,
  Bromley, and Rollin}}]{mitroy09b}
\bibinfo{author}{\bibfnamefont{J.}~\bibnamefont{Mitroy}},
  \bibinfo{author}{\bibfnamefont{J.~Y.} \bibnamefont{Zhang}},
  \bibinfo{author}{\bibfnamefont{M.~W.~J.} \bibnamefont{Bromley}},
  \bibnamefont{and} \bibinfo{author}{\bibfnamefont{K.~G.}
  \bibnamefont{Rollin}}, \bibinfo{journal}{Eur.~Phys.~J.~D}
  \textbf{\bibinfo{volume}{53}}, \bibinfo{pages}{15} (\bibinfo{year}{2009}).

\bibitem[{\citenamefont{{Grant} and {Quiney}}(2000)}]{grant00a}
\bibinfo{author}{\bibfnamefont{I.~P.} \bibnamefont{{Grant}}} \bibnamefont{and}
  \bibinfo{author}{\bibfnamefont{H.~M.} \bibnamefont{{Quiney}}},
  \bibinfo{journal}{\pra} \textbf{\bibinfo{volume}{62}}, \bibinfo{eid}{022508}
  (\bibinfo{year}{2000}).

\bibitem[{\citenamefont{Grant}(2007)}]{grant07a}
\bibinfo{author}{\bibfnamefont{I.~P.} \bibnamefont{Grant}},
  \emph{\bibinfo{title}{Relativistic Quantum Theory of Atoms and Molecules
  Theory and Computation}} (\bibinfo{publisher}{Springer},
  \bibinfo{address}{New York}, \bibinfo{year}{2007}).

\bibitem[{\citenamefont{{Safronova} et~al.}(2013)\citenamefont{{Safronova},
  {Safronova}, and {Clark}}}]{safronova13a}
\bibinfo{author}{\bibfnamefont{M.~S.} \bibnamefont{{Safronova}}},
  \bibinfo{author}{\bibfnamefont{U.~I.} \bibnamefont{{Safronova}}},
  \bibnamefont{and} \bibinfo{author}{\bibfnamefont{C.~W.}
  \bibnamefont{{Clark}}}, \bibinfo{journal}{ArXiv e-prints}
  (\bibinfo{year}{2013}), \eprint{1301.3181}.

\bibitem[{\citenamefont{Safronova and Johnson}(2008)}]{safronova08a}
\bibinfo{author}{\bibfnamefont{M.~S.} \bibnamefont{Safronova}}
  \bibnamefont{and} \bibinfo{author}{\bibfnamefont{W.~R.}
  \bibnamefont{Johnson}}, \bibinfo{journal}{Adv.~At.~Mol.~Opt.~Phys.}
  \textbf{\bibinfo{volume}{55}}, \bibinfo{pages}{191} (\bibinfo{year}{2008}).

\bibitem[{\citenamefont{Kramida et~al.}(2012)\citenamefont{Kramida, Ralchenko,
  Reader, and {NIST ASD Team}}}]{nistasd500}
\bibinfo{author}{\bibfnamefont{A.}~\bibnamefont{Kramida}},
  \bibinfo{author}{\bibfnamefont{Y.}~\bibnamefont{Ralchenko}},
  \bibinfo{author}{\bibfnamefont{J.}~\bibnamefont{Reader}}, \bibnamefont{and}
  \bibinfo{author}{\bibnamefont{{NIST ASD Team}}}, \emph{\bibinfo{title}{{NIST
  Atomic Spectra Database (version 5.0.0)}}} (\bibinfo{year}{2012}),
  \urlprefix\url{http://physics.nist.gov/asd}.

\bibitem[{\citenamefont{Hameed}(1972)}]{hameed72a}
\bibinfo{author}{\bibfnamefont{S.}~\bibnamefont{Hameed}},
  \bibinfo{journal}{J.~Phys.~B} \textbf{\bibinfo{volume}{5}},
  \bibinfo{pages}{746} (\bibinfo{year}{1972}).

\bibitem[{\citenamefont{Hameed et~al.}(1968)\citenamefont{Hameed, Herzenberg,
  and James}}]{hameed68a}
\bibinfo{author}{\bibfnamefont{S.}~\bibnamefont{Hameed}},
  \bibinfo{author}{\bibfnamefont{A.}~\bibnamefont{Herzenberg}},
  \bibnamefont{and} \bibinfo{author}{\bibfnamefont{M.~G.} \bibnamefont{James}},
  \bibinfo{journal}{J.~Phys.~B} \textbf{\bibinfo{volume}{1}},
  \bibinfo{pages}{822} (\bibinfo{year}{1968}).

\bibitem[{\citenamefont{Safronova}(2013)}]{safronova13c}
\bibinfo{author}{\bibfnamefont{M.~S.} \bibnamefont{Safronova}}
  (\bibinfo{year}{2013}), \bibinfo{note}{(private communication)}.

\bibitem[{\citenamefont{{Safronova} and {Safronova}}(2008)}]{safronova08b}
\bibinfo{author}{\bibfnamefont{U.~I.} \bibnamefont{{Safronova}}}
  \bibnamefont{and} \bibinfo{author}{\bibfnamefont{M.~S.}
  \bibnamefont{{Safronova}}}, \bibinfo{journal}{Phys.~Rev.~A}
  \textbf{\bibinfo{volume}{78}}, \bibinfo{pages}{052504}
  (\bibinfo{year}{2008}).

\bibitem[{\citenamefont{Mitroy and Bromley}(2005)}]{mitroy05b}
\bibinfo{author}{\bibfnamefont{J.}~\bibnamefont{Mitroy}} \bibnamefont{and}
  \bibinfo{author}{\bibfnamefont{M.~W.~J.} \bibnamefont{Bromley}},
  \bibinfo{journal}{Phys.~Rev.~A} \textbf{\bibinfo{volume}{71}},
  \bibinfo{pages}{019902(E),019903(E)} (\bibinfo{year}{2005}).

\bibitem[{\citenamefont{Mitroy and Bromley}(2003{\natexlab{b}})}]{mitroy03g}
\bibinfo{author}{\bibfnamefont{J.}~\bibnamefont{Mitroy}} \bibnamefont{and}
  \bibinfo{author}{\bibfnamefont{M.~W.~J.} \bibnamefont{Bromley}},
  \bibinfo{journal}{Phys.~Rev.~A} \textbf{\bibinfo{volume}{68}},
  \bibinfo{pages}{062710} (\bibinfo{year}{2003}{\natexlab{b}}).

\bibitem[{\citenamefont{Mitroy and Zhang}(2007)}]{mitroy07d}
\bibinfo{author}{\bibfnamefont{J.}~\bibnamefont{Mitroy}} \bibnamefont{and}
  \bibinfo{author}{\bibfnamefont{J.~Y.} \bibnamefont{Zhang}},
  \bibinfo{journal}{Phys.~Rev.~A} \textbf{\bibinfo{volume}{76}},
  \bibinfo{pages}{032706} (\bibinfo{year}{2007}).

\bibitem[{\citenamefont{Safronova et~al.}(1999)\citenamefont{Safronova,
  Johnson, and Derevianko}}]{safronova99a}
\bibinfo{author}{\bibfnamefont{M.~S.} \bibnamefont{Safronova}},
  \bibinfo{author}{\bibfnamefont{W.~R.} \bibnamefont{Johnson}},
  \bibnamefont{and}
  \bibinfo{author}{\bibfnamefont{A.}~\bibnamefont{Derevianko}},
  \bibinfo{journal}{Phys.~Rev.~A} \textbf{\bibinfo{volume}{60}},
  \bibinfo{pages}{4476} (\bibinfo{year}{1999}).

\bibitem[{\citenamefont{Volz and Schmoranzer}(1996)}]{volz96a}
\bibinfo{author}{\bibfnamefont{U.}~\bibnamefont{Volz}} \bibnamefont{and}
  \bibinfo{author}{\bibfnamefont{H.}~\bibnamefont{Schmoranzer}},
  \bibinfo{journal}{Phys.~Scr.} \textbf{\bibinfo{volume}{T65}},
  \bibinfo{pages}{48} (\bibinfo{year}{1996}).

\bibitem[{\citenamefont{{Falke} et~al.}(2006)\citenamefont{{Falke}, {Sherstov},
  {Tiemann}, and {Lisdat}}}]{falke06a}
\bibinfo{author}{\bibfnamefont{S.}~\bibnamefont{{Falke}}},
  \bibinfo{author}{\bibfnamefont{I.}~\bibnamefont{{Sherstov}}},
  \bibinfo{author}{\bibfnamefont{E.}~\bibnamefont{{Tiemann}}},
  \bibnamefont{and} \bibinfo{author}{\bibfnamefont{C.}~\bibnamefont{{Lisdat}}},
  \bibinfo{journal}{\jcp} \textbf{\bibinfo{volume}{125}},
  \bibinfo{pages}{224303} (\bibinfo{year}{2006}).

\bibitem[{\citenamefont{{Shabanova, L N and Khlustalov, A
  N}}(1984)}]{shabanova84a}
\bibinfo{author}{\bibnamefont{{Shabanova, L N and Khlustalov, A N}}},
  \bibinfo{journal}{Opt.~Spectrosc.} \textbf{\bibinfo{volume}{59}},
  \bibinfo{pages}{123} (\bibinfo{year}{1984}), \bibinfo{note}{optika i
  Spectrosk. \textbf{53} 600 (1982)}.

\bibitem[{\citenamefont{{Migdalek} and {Kim}}(1998)}]{migdalek98a}
\bibinfo{author}{\bibfnamefont{J.}~\bibnamefont{{Migdalek}}} \bibnamefont{and}
  \bibinfo{author}{\bibfnamefont{Y.-K.} \bibnamefont{{Kim}}},
  \bibinfo{journal}{J.~Phys.~B} \textbf{\bibinfo{volume}{31}},
  \bibinfo{pages}{1947} (\bibinfo{year}{1998}).

\bibitem[{\citenamefont{Holmgren et~al.}(2010)\citenamefont{Holmgren, Revelle,
  Lonij, and Cronin}}]{holmgren10a}
\bibinfo{author}{\bibfnamefont{W.~F.} \bibnamefont{Holmgren}},
  \bibinfo{author}{\bibfnamefont{M.~C.} \bibnamefont{Revelle}},
  \bibinfo{author}{\bibfnamefont{V.~P.~A.} \bibnamefont{Lonij}},
  \bibnamefont{and} \bibinfo{author}{\bibfnamefont{A.~D.}
  \bibnamefont{Cronin}}, \bibinfo{journal}{Phys.~Rev.~A}
  \textbf{\bibinfo{volume}{81}}, \bibinfo{pages}{053607}
  (\bibinfo{year}{2010}).

\bibitem[{\citenamefont{Derevianko et~al.}(1999)\citenamefont{Derevianko,
  Johnson, Safronova, and Babb}}]{derevianko99a}
\bibinfo{author}{\bibfnamefont{A.}~\bibnamefont{Derevianko}},
  \bibinfo{author}{\bibfnamefont{W.~R.} \bibnamefont{Johnson}},
  \bibinfo{author}{\bibfnamefont{M.~S.} \bibnamefont{Safronova}},
  \bibnamefont{and} \bibinfo{author}{\bibfnamefont{J.~F.} \bibnamefont{Babb}},
  \bibinfo{journal}{Phys.~Rev.~Lett.} \textbf{\bibinfo{volume}{82}},
  \bibinfo{pages}{3589} (\bibinfo{year}{1999}).

\bibitem[{\citenamefont{{Lim} et~al.}(1999)\citenamefont{{Lim}, {Pernpointner},
  {Seth}, {Laerdahl}, {Schwerdtfeger}, {Neogrady}, and {Urban}}}]{lim99a}
\bibinfo{author}{\bibfnamefont{I.~S.} \bibnamefont{{Lim}}},
  \bibinfo{author}{\bibfnamefont{M.}~\bibnamefont{{Pernpointner}}},
  \bibinfo{author}{\bibfnamefont{M.}~\bibnamefont{{Seth}}},
  \bibinfo{author}{\bibfnamefont{J.~K.} \bibnamefont{{Laerdahl}}},
  \bibinfo{author}{\bibfnamefont{P.}~\bibnamefont{{Schwerdtfeger}}},
  \bibinfo{author}{\bibfnamefont{P.}~\bibnamefont{{Neogrady}}},
  \bibnamefont{and} \bibinfo{author}{\bibfnamefont{M.}~\bibnamefont{{Urban}}},
  \bibinfo{journal}{Phys.~Rev.~A} \textbf{\bibinfo{volume}{60}},
  \bibinfo{pages}{2822} (\bibinfo{year}{1999}).

\bibitem[{\citenamefont{Molof et~al.}(1974)\citenamefont{Molof, Schwartz,
  Miller, and Bederson}}]{molof74a}
\bibinfo{author}{\bibfnamefont{R.~W.} \bibnamefont{Molof}},
  \bibinfo{author}{\bibfnamefont{H.~L.} \bibnamefont{Schwartz}},
  \bibinfo{author}{\bibfnamefont{T.~M.} \bibnamefont{Miller}},
  \bibnamefont{and} \bibinfo{author}{\bibfnamefont{B.}~\bibnamefont{Bederson}},
  \bibinfo{journal}{Phys.~Rev.~A} \textbf{\bibinfo{volume}{10}},
  \bibinfo{pages}{1131} (\bibinfo{year}{1974}).

\bibitem[{\citenamefont{{Margoliash} and {Meath}}(1978)}]{margoliash78a}
\bibinfo{author}{\bibfnamefont{D.~J.} \bibnamefont{{Margoliash}}}
  \bibnamefont{and} \bibinfo{author}{\bibfnamefont{W.~J.}
  \bibnamefont{{Meath}}}, \bibinfo{journal}{J.~Chem.~Phys.}
  \textbf{\bibinfo{volume}{68}}, \bibinfo{pages}{1426} (\bibinfo{year}{1978}).

\bibitem[{\citenamefont{Kumar and Meath}(1985)}]{kumar85a}
\bibinfo{author}{\bibfnamefont{A.}~\bibnamefont{Kumar}} \bibnamefont{and}
  \bibinfo{author}{\bibfnamefont{W.~J.} \bibnamefont{Meath}},
  \bibinfo{journal}{Mol. Phys.} \textbf{\bibinfo{volume}{54}},
  \bibinfo{pages}{823} (\bibinfo{year}{1985}).

\bibitem[{\citenamefont{{\"{O}}pik}(1967)}]{opik67a}
\bibinfo{author}{\bibfnamefont{U.}~\bibnamefont{{\"{O}}pik}},
  \bibinfo{journal}{Proc.~Phys.~Soc.~London} \textbf{\bibinfo{volume}{92}},
  \bibinfo{pages}{566} (\bibinfo{year}{1967}).

\bibitem[{\citenamefont{Mitroy}(1993)}]{mitroy93a}
\bibinfo{author}{\bibfnamefont{J.}~\bibnamefont{Mitroy}},
  \bibinfo{journal}{J.~Phys.~B} \textbf{\bibinfo{volume}{26}},
  \bibinfo{pages}{2201} (\bibinfo{year}{1993}).

\bibitem[{\citenamefont{{Sansonetti} et~al.}(2011)\citenamefont{{Sansonetti},
  {Simien}, {Gillaspy}, {Tan}, {Brewer}, {Brown}, {Wu}, and
  {Porto}}}]{sansonetti11a}
\bibinfo{author}{\bibfnamefont{C.~J.} \bibnamefont{{Sansonetti}}},
  \bibinfo{author}{\bibfnamefont{C.~E.} \bibnamefont{{Simien}}},
  \bibinfo{author}{\bibfnamefont{J.~D.} \bibnamefont{{Gillaspy}}},
  \bibinfo{author}{\bibfnamefont{J.~N.} \bibnamefont{{Tan}}},
  \bibinfo{author}{\bibfnamefont{S.~M.} \bibnamefont{{Brewer}}},
  \bibinfo{author}{\bibfnamefont{R.~C.} \bibnamefont{{Brown}}},
  \bibinfo{author}{\bibfnamefont{S.}~\bibnamefont{{Wu}}}, \bibnamefont{and}
  \bibinfo{author}{\bibfnamefont{J.~V.} \bibnamefont{{Porto}}},
  \bibinfo{journal}{Phys.~Rev.~Lett.} \textbf{\bibinfo{volume}{107}},
  \bibinfo{eid}{023001} (\bibinfo{year}{2011}).

\bibitem[{\citenamefont{Radziemski et~al.}(1995)\citenamefont{Radziemski,
  Engleman, and Brault}}]{radziemski95a}
\bibinfo{author}{\bibfnamefont{L.~J.} \bibnamefont{Radziemski}},
  \bibinfo{author}{\bibfnamefont{R.}~\bibnamefont{Engleman}}, \bibnamefont{and}
  \bibinfo{author}{\bibfnamefont{J.~W.} \bibnamefont{Brault}},
  \bibinfo{journal}{Phys.~Rev.~A} \textbf{\bibinfo{volume}{52}},
  \bibinfo{pages}{4462} (\bibinfo{year}{1995}).

\bibitem[{\citenamefont{Risberg}(1956)}]{risberg56b}
\bibinfo{author}{\bibfnamefont{V.}~\bibnamefont{Risberg}},
  \bibinfo{journal}{Ark.~Phys} \textbf{\bibinfo{volume}{10}},
  \bibinfo{pages}{583} (\bibinfo{year}{1956}).

\bibitem[{\citenamefont{{Zatsarinny} and {Tayal}}(2010)}]{zatsarinny10a}
\bibinfo{author}{\bibfnamefont{O.}~\bibnamefont{{Zatsarinny}}}
  \bibnamefont{and} \bibinfo{author}{\bibfnamefont{S.~S.}
  \bibnamefont{{Tayal}}}, \bibinfo{journal}{\pra}
  \textbf{\bibinfo{volume}{81}}, \bibinfo{pages}{043423}
  (\bibinfo{year}{2010}).

\end{thebibliography}

\end{document}